\documentclass[12pt]{article}
\usepackage{graphicx,amsmath,amsthm,amscd,amsfonts}

\usepackage{amssymb}
\input xy
\xyoption{all} 

\begin{document}

\vspace*{0.3in}

\begin{center}

{\large\bf Predictions for Gromov-Witten invariants}

{\large\bf  of noncommutative resolutions}

\vspace{0.5in}

Eric Sharpe

Physics Department\\
Robeson Hall, 0435\\
Virginia Tech \\
Blacksburg, VA  24061

{\tt ersharpe@vt.edu}

\end{center}

In this paper, we apply recent methods of localized GLSM's to make
predictions for Gromov-Witten invariants of noncommutative resolutions,
as defined by {\it e.g.} Kontsevich, and use those predictions to
examine the connectivity of the SCFT moduli space through complex
structure deformations.  
Noncommutative spaces, in the
present sense, are defined by their sheaves, their B-branes.
Examples of abstract CFT's whose B-branes correspond with those
defining noncommutative spaces arose in examples of abelian GLSM's describing
branched double covers, in which the double cover structure arises
nonperturbatively.  
This note will examine the GLSM for ${\mathbb P}^7[2,2,2,2]$,
which realizes this phenomenon.
Its Landau-Ginzburg
point is a noncommutative resolution of a (singular) branched double cover
of ${\mathbb P}^3$.  Regardless of the complex structure of the
large-radius ${\mathbb P}^7[2,2,2,2]$, the Landau-Ginzburg point is
always a noncommutative resolution of a singular space, which begs the
question of whether the noncommutative resolution is connected in
SCFT moduli space by a complex structure deformation to a smooth branched
double cover.  Using recent localization techniques, we make a prediction
for the Gromov-Witten invariants of the noncommutative resolution,
and find that they do not match those of a smooth branched double cover,
telling us that these abstract CFT's are not continuously connected to
sigma models on smooth branched double covers through complex structure
deformations.

\begin{flushleft}
December 2012
\end{flushleft}

\newpage

\tableofcontents

\section{Introduction}

This short note describes a prediction for Gromov-Witten invariants
of an abstract CFT, interpreted in \cite{cdhps} as a 
`noncommutative resolution' (in Kontsevich's sense) of a singular branched
double cover, and its application to determine connectivity of the
moduli space of such CFT's.

The noncommutative resolutions in question are defined mathematically
just by their sheaf theory.  That by itself does not suffice to define
a CFT; instead, the work \cite{cdhps} found examples of gauged linear
sigma models (GLSMs) which, at special points, RG flowed to abstract
CFT's with the same B-branes as those defined by noncommutative
resolutions.  For that reason, those abstract CFT's were identified
with physical realizations of the noncommutative resolutions.
(This also means that this notion of noncommutative geometry has a different
physical realization than previous realizations of noncommutative
geometry in physics such as \cite{swnc,rwnc}.)

One purpose of this note is simply to make a prediction for Gromov-Witten
invariants of the noncommutative resolutions appearing at the 
Landau-Ginzburg point of the GLSM for ${\mathbb P}^7[2,2,2,2]$.
Now, it is not clear that a noncommutative resolution, as defined
mathematically, should admit a notion of Gromov-Witten invariants,
as they need not come with {\it e.g.} symplectic structures, but here
we have a full-fledged CFT whose B-branes coincide with those of the
noncommutative resolution, so one can hope to apply the extra structure
implicit in the CFT.  That said,
since they are abstract CFT's, which might not have an analogue of
a large-radius limit (as suggested by monodromy computations in
\cite{cdhps}), it is still not in principle obvious that
Gromov-Witten invariants can be defined.  Nevertheless, using the methods
of \cite{jklmr2}, we are able to compute Gromov-Witten invariants of these
theories.

Another application of those Gromov-Witten invariants is to test the
connectivity of the SCFT moduli space.
If the noncommutative resolutions in question are connected in
SCFT moduli space to smooth branched double covers via a complex structure
deformation, then A model correlation functions and Gromov-Witten invariants
of the noncommutative resolutions should match those of
smooth branched double covers.  
However, it is not clear that the moduli spaces need be so
connected.  For example, the GLSM for ${\mathbb P}^7[2,2,2,2]$ always
generates noncommutative resolutions at the Landau-Ginzburg point, 
never smooth branched double covers, and conversely starting from a smooth
branched double cover it is not clear physically what marginal operator
would be responsible for generating a noncommutative resolution.
As a result, it is not {\it a priori} obvious that the noncommutative
resolutions and smooth branched double covers necessarily lie on the same SCFT
moduli space, and one of the points of this paper is to examine this
question.

When we compute Gromov-Witten invariants, we find that the curve counting
in the noncommutative resolution is different from that in a
corresponding smooth branched double cover, hence, as we suggested above,
the two components of the SCFT moduli space cannot be connected through a 
complex structure deformation.  Apparently the noncommutative
resolutions arising at the Landau-Ginzburg point of the GLSM for
${\mathbb P}^7[2,2,2,2]$ are examples of `frozen' singularities, in the
sense that the singularity and noncommutative resolution cannot be
deformed away.

We begin in section~\ref{sect:glsm:nc} by reviewing pertinent aspects
of the analysis of the GLSM for ${\mathbb P}^7[2,2,2,2]$, and the
appearance of noncommutative resolutions.  As a warm-up,
in section~\ref{sect:lr} we compute Gromov-Witten invariants for 
${\mathbb P}^7[2,2,2,2]$ using the methods of \cite{jklmr2}.
These Gromov-Witten invariants are known, and we recover the standard
result.  Finally, in section~\ref{sect:nc} we use the same methods to
compute Gromov-Witten invariants at the Landau-Ginzburg point of the
same model, which is interpreted as a noncommutative resolution of a 
singular branched double cover.

\section{Review of pertinent GLSM's}  
\label{sect:glsm:nc}

Up until several years ago, it was thought that all gauged
linear sigma models:
\begin{itemize}
\item Can only describe geometries that are global complete
intersections,
\item Those geometries can be realized physically only as the critical
locus of a superpotential, and
\item All geometric phases of GLSM's are birational to one another.
\end{itemize}
The papers \cite{horitong,ron-sharpe,cdhps} found counterexamples to all
of the statements above, in both nonabelian \cite{horitong,ron-sharpe} and
abelian \cite{cdhps} GLSM's.  The nonabelian examples produced
Pfaffian and determinantal varieties, realized by strongly-coupled
gauge theoretic
effects (and more recently, realized perturbatively \cite{hori2,jklmr1}),
and the abelian examples gave (noncommutative resolutions of) branched
double covers, also realized via nonperturbative effects.  
These branched-double-cover structures have been independently
checked in {\it e.g.} 
\cite{hori2}[section 6.5] using gauge-theoretic dualities and
\cite{ed-nick-me} using an analysis of D-brane probes and
matrix factorizations.  Although the different phases in these examples
are not birational, they are instead related by `homological projective
duality' \cite{kuz1,kuz2,kuz3}, and more recent work \cite{bfk} strongly
supports the assertion that all phases of GLSM's are related by
homological projective duality.  See \cite{rev1,rev2,rev3} for reviews
of the abelian GLSM examples and branched double covers.

The `noncommutative resolutions' referred to above, are the focus of
this article.  Mathematically, these are certain generalized notions of
spaces defined by their sheaf theory, broadly speaking; see
\cite{kuz2} for the specific pertinent noncommutative resolutions,
and {\it e.g.} \cite{vdb1,vdb2,vdb3,blb,dao,bl} for closely related
material.  To define a CFT associated to a noncommutative resolution,
one needs more data than just a set of sheaves.  What was uncovered
in \cite{cdhps} are a set of abstract CFT's, which look `mostly' like
branched double covers, away from singularities, and which everywhere
possess B-branes matching those defining the noncommutative resolution.
For that reason, matching B-branes, the abstract CFT's were identified
with a physical realization of
noncommutative resolutions, a result which also matched a mathematical
prediction of homological projective duality.

Let us quickly review the structure of the abelian GLSM's in which
branched double covers arose, beginning with an example in which
no noncommutative resolution was present.
The simplest example discussed in \cite{cdhps} was the
GLSM for the complete intersection Calabi-Yau ${\mathbb P}^3[2,2]$.
The superpotential for this theory is of the form
\begin{displaymath}
W \: = \: \sum_a p_a G_a(\phi) \: = \:
\sum_{ij} \phi_i \phi_j A^{ij}(p),
\end{displaymath}
where the $\phi$'s act as homogeneous coordinates on ${\mathbb P}^3$,
the $G_a$'s are the two quadrics, and $A^{ij}$ is a symmetric $4 \times 4$
matrix with entries linear in the $p$'s, determined by the $G_a$'s.

At the Landau-Ginzburg point of this theory, where the $p_a$ are not
all zero, the superpotential acts as a mass matrix for $\phi$'s.
Naively, this is problematic:  we are left with a theory containing
only $p$'s, which looks like a sigma model on ${\mathbb P}^1$, which cannot
possibly be Calabi-Yau.  However, a closer analysis reveals subtleties.
First, since the $p$'s are charge 2, there is a trivially-acting
${\mathbb Z}_2$ here (technically, a ${\mathbb Z}_2$ gerbe structure),
which physics interprets \cite{hhpsa} as a double cover.
Second, the mass matrix $A^{ij}(p)$ has zero eigenvalues along the 
degree four hypersurface $\{ \det A = 0\}$.  
With a bit of further analysis discussed in
\cite{cdhps}, one argues that this flows in the IR to a nonlinear sigma
model on 
a branched
double cover of ${\mathbb P}^1$, branched over a degree four
hypersurface --
an example of a Calabi-Yau.  In fact, both ${\mathbb P}^3[2,2]$
and the branched double cover are elliptic curves.

Analogous analyses apply to many other examples.  The next simplest
involves the GLSM for ${\mathbb P}^5[2,2,2]$, which is a K3 surface.
Its Landau-Ginzburg point is interpreted as a branched double cover
of ${\mathbb P}^2$, branched over a degree six locus, which is another
K3.

The case after that is more interesting.  The Landau-Ginzburg point of the
GLSM for ${\mathbb P}^7[2,2,2,2]$ is, naively, a branched double cover of
${\mathbb P}^3$.  However, mathematically that branched double cover is
always singular, and yet the GLSM behaves as if it is describing a smooth
space.  The resolution described in \cite{cdhps} is that the GLSM is
instead describing a `noncommutative resolution' of the singular
branched double cover.  That structure can be seen most directly in
matrix factorizations in a Landau-Ginzburg model intermediate in RG flow.
The noncommutative resolution is defined by its sheaf theory,
and specifically, its sheaves are all sheaves of $B$-modules
over ${\mathbb P}^3$ (equivalently, sheaves of modules over Azumaya algebras
over the branched double cover), for $B$ a sheaf of even parts of Clifford
algebras defined by the GLSM superpotential.  
Matrix factorizations in the Landau-Ginzburg model automatically have this
structure, hence we can identify the Landau-Ginzburg
model with a physical realization of the noncommutative resolution.

In this paper we shall apply the ideas of
\cite{jklmr2,gl} to predict the Gromov-Witten invariants of noncommutative
spaces, such as those discussed in \cite{cdhps}.

\section{${\mathbb P}^7[2,2,2,2]$:  large-radius analysis}
\label{sect:lr}

Let us consider the GLSM for ${\mathbb P}^7[2,2,2,2]$.
This GLSM has two sets of fields:  $\Phi_i$, $i\in\{1, \cdots, 8\}$,
with gauge $U(1)$ charge $1$ and $U(1)_V$ charge $2\mathsf{q}$,
and $P_a$, $a \in \{1,\cdots,4\}$, with gauge $U(1)$ charge $-2$
and $U(1)_V$ charge $2-4\mathsf{q}$.
Following the discussion in \cite{jklmr2}, the partition function\footnote{
In principle, one expects that
the expression above will pick up a phase as one wanders
around on the SCFT moduli space and the phases of the
localizing supercharges vary \cite{dist-trieste}, hence the
particular expression above corresponds to a particular choice of
normalization. 
} for
this theory for $r \gg 0$ is \cite{benini1,doroud1}
\begin{displaymath}
Z_{nlsm} \: = \: \sum_{m \in {\mathbb Z}} e^{-i \theta m}
\int_{-\infty}^{\infty} \frac{d \sigma}{2\pi} e^{-4 \pi i r \sigma}
(Z_{\Phi})^8 (Z_P)^4,
\end{displaymath}
where
\begin{displaymath}
Z_{\Phi} \: = \: \frac{ \Gamma( \mathsf{q} -  i \sigma -  m/2) }{ 
\Gamma( 1 - \mathsf{q} +  i \sigma -  m/2) } , \: \: \:
Z_P \: = \: \frac{ \Gamma( (1-2\mathsf{q}) + 2 i \sigma + 2 m/2) }{
\Gamma( 2\mathsf{q} - 2 i \sigma + 2 m/2) }.
\end{displaymath}

As in \cite{jklmr2}, define $\tau = \mathsf{q} - i \sigma$, 
so that the partition
function above becomes
\begin{equation}  \label{start-part-fn}
Z_{nlsm} \: = \: e^{- 4 \pi \mathsf{q} r}
\sum_{m \in {\mathbb Z}} e^{-i \theta m}
\int_{\mathsf{q}-i\infty}^{\mathsf{q}+i\infty} \frac{d \tau}{2\pi i} 
e^{4 \pi  r \tau}
\left(
\frac{ \Gamma( \tau - m/2 ) }{ \Gamma( 1- \tau - m/2) }
\right)^8
\left(
\frac{ \Gamma( 1 - 2 \tau + m ) }{ \Gamma(2 \tau + m ) }
\right)^4.
\end{equation}

We shall first consider the $r \gg 0$ region in the K\"ahler moduli space,
and verify that the method of \cite{jklmr2} recovers the Gromov-Witten
invariants of ${\mathbb P}^7[2,2,2,2]$.  Then, we shall apply the same
method to the $r \ll 0$ phase to make a prediction for Gromov-Witten invariants
of the noncommutative resolution of the branched double cover.

When $r \gg 0$, we can close the contour above on the left half-plane.
Assume $0 < \mathsf{q} < 1/2$.

Let us find poles that will contribute to the contour integral above.

First, we claim that there will be no net contribution from $Z_P$, in the
sense that all poles in the numerator are cancelled out by corresponding
poles in the denominator.  (The rest may have zeroes at some of the
zeroes of $Z_{\Phi}$, on the other hand.)
First, note $\Gamma(1-2\tau + m)$ has poles at
\begin{displaymath}
\tau \: = \: \frac{1}{2}\left( m \: + \: 1 \: + \: k \right) 
\end{displaymath}
for $k \geq 0$, and these will lie inside the contour when
$k + 1 < - m$.  Since $k \geq 0$, this can only happen if $m < 0$,
in which case, $0 \leq k < - (m+1)$.
Similarly, $\Gamma(2\tau + m)^{-1}$ has zeroes when
\begin{displaymath}
\tau \: = \: - \frac{1}{2}\left( m \: + \: k \right)
\end{displaymath}
for $k \geq 0$, and these will lie inside the contour when
$-m - k \leq 0$, {\it i.e.} $k \geq {\rm max}\{0, - m \}$.
In particular, any pole of $\Gamma(1-2\tau +m)$, defined by some $k$,
is matched by a zero of $\Gamma(2 \tau + m)^{-1}$ at
\begin{displaymath}
\tau \: = \: - \frac{1}{2} \left( m \: + \: \left( - 2m - 1 - k \right) 
\right).
\end{displaymath}
As a consistency check, note that since $k < -m-1$, 
\begin{displaymath}
-2m -1 -k \: > \: -2m -1 +m +1 \: = \: -m
\end{displaymath}
hence (as $m < 0$) $-2m-1-k$ is in the right range to define a zero of
$\Gamma(2\tau + m)^{-1}$.

Now that we have established that $Z_P$ will not contribute to the
pole count, let us turn to $Z_{\Phi}$.
The numerator
$\Gamma(\tau - m/2)$ has poles at
\begin{displaymath}
\tau \: = \: m/2 \: - \: k
\end{displaymath}
for $k \geq 0$, and these will lie inside the integration contour when
$k \geq m/2$.  

In the case of $Z_P$, all poles were matched by zeroes, but for
$Z_{\Phi}$, only some of the poles will be matched by zeroes.
The remaining unmatched poles will be counted by $k \geq {\rm max}\{0, m\}$.
We can see this as follows.
The zeroes of $\Gamma(1 - \tau - m/2)^{-1}$ are located at
\begin{displaymath}
\tau \: = \: 1 \: - \: m/2 \: + \: k_1
\end{displaymath}
for $k_1 \geq 0$.
A zero coincides with a pole when
\begin{displaymath}
1 \: - \: m/2 \: + \: k_1 \: = \:
m/2 \: - \: k
\end{displaymath}
or equivalently
\begin{displaymath}
k_1 \: = \: m \: - \: k \: - \: 1 \: \geq \: 0,
\end{displaymath}
which requires $k \leq m-1$.
Thus, we see that if $0 \leq k \leq m-1$, then the corresponding
pole in $Z_{\Phi}$ is cancelled by a zero, so we only have 
(unmatched) poles in $Z_{\Phi}$ for
$k \geq {\rm max}\{0, m \}$, or equivalently $m \leq k$.

Next, let us evaluate the contour integral above.
First, let us rewrite it as a sum of residues: 
\begin{displaymath}
Z_{nlsm} \: = \:  \sum_{k=0}^{\infty} \sum_{m \leq k}
e^{-i m \theta} e^{-4 \pi \mathsf{q} r}
{\rm Res}_{\tau = m/2 - k} 
\left\{ e^{4 \pi r \tau} 
\left(
\frac{ \Gamma( \tau - m/2 ) }{ \Gamma( 1- \tau - m/2) }
\right)^8
\left(
\frac{ \Gamma( 1 - 2 \tau + m ) }{ \Gamma(2 \tau + m ) }
\right)^4
\right\}.
\end{displaymath}

Now, let us simplify this expression, following \cite{jklmr2}[appendix A].  
First, define $\ell$ by $m=k-\ell$,
so that the expression above becomes
\begin{displaymath}
\sum_{k=0}^{\infty} \sum_{\ell=0}^{\infty}
e^{-i (k-\ell) \theta} e^{-4 \pi \mathsf{q} r} \oint \frac{d \epsilon}{2 \pi i}
\left\{
e^{4 \pi r( - \ell/2 - k/2 + \epsilon)}
\left( \frac{ \Gamma(\epsilon - k) }{\Gamma(1 + \ell - \epsilon) } \right)^8
\left( \frac{\Gamma(1 + 2k - 2 \epsilon) }{\Gamma(-2\ell - 2 \epsilon) }
\right)^4
\right\}.
\end{displaymath}
Then, define $z = \exp(- 2 \pi r + i \theta)$ and use the identity
\begin{displaymath}
\Gamma(x) \: = \: \frac{\pi}{\sin \pi x} \frac{1}{\Gamma(1 - x)}
\end{displaymath}
to write
\begin{eqnarray*}
Z_{nlsm} & = & \oint \frac{d \epsilon}{2 \pi i}
\sum_{k=0}^{\infty} \sum_{\ell=0}^{\infty}
z^{\ell} \overline{z}^k (z \overline{z})^{\mathsf{q} - \epsilon}
\pi^{4} \frac{ ( \sin \pi (-2 \ell + 2 \epsilon) )^4 }{
( \sin \pi (\epsilon - k) )^8 }
\frac{\Gamma(1 + 2k - 2 \epsilon)^4 }{\Gamma(1 + k - \epsilon )^8 }
\frac{ \Gamma(1 + 2 \ell - 2 \epsilon)^4 }{\Gamma(1 + \ell - \epsilon)^8 },
\\
& = & \oint \frac{d \epsilon}{2 \pi i} (z \overline{z})^{\mathsf{q} - \epsilon}
\pi^4 \frac{ ( \sin 2 \pi \epsilon )^4 }{ (\sin \pi \epsilon)^8 }
\left|
\sum_{k=0}^{\infty} z^k 
\frac{\Gamma(1 + 2k - 2 \epsilon)^4 }{ \Gamma(1 + k - \epsilon)^8 } 
\right|^2,
\end{eqnarray*}
where the complex conjugation acts only on $z$, not $\epsilon$.

To evaluate this, first define
\begin{displaymath}
f(\epsilon) \: = \:
\left|
\sum_{k=0}^{\infty} z^k 
\frac{\Gamma(1 + 2k - 2 \epsilon)^4 }{ \Gamma(1 + k - \epsilon)^8 } 
\right|^2.
\end{displaymath}
Then, it is straightforward to show that
\begin{eqnarray*}
\lefteqn{
Z_{nlsm} \: = \:
\frac{8}{3} (z \overline{z})^{\mathsf{q}} \left[ - \ln(z \overline{z})^3 f(0)
\: - \: 8 \pi^2 f'(0) \: + \: 3 \ln(z \overline{z})^2 f'(0)
\right. } \\
& & \hspace*{1in} \left.
\: + \: \ln(z \overline{z})\left( 8 \pi^2 f(0) - 3 f''(0) \right)
\: + \: f^{(3)}(0) \right].
\end{eqnarray*}

Now, in principle, a normalized $Z_{nlsm}$ in a one-parameter model
such as this should match 
\cite{jklmr2}[equ'n (2.19)]
\begin{eqnarray}
\exp(-K) & = & - \frac{i}{6} \kappa (t - \overline{t})^3 \: + \:
\frac{\zeta(3)}{4\pi^3} \chi(X) 
 \: + \: \frac{2i}{(2\pi i)^3} \sum_n N_n \left(
{\rm Li}_3(q^n) + {\rm Li}_3( \overline{q}^n ) \right) 
\nonumber \\
& & \hspace*{0.5in}
\: - \: \frac{i}{(2\pi i)^2} \sum_n N_n\left(
{\rm Li}_2(q^n) + {\rm Li}_2( \overline{q}^n) \right)
n (t - \overline{t}),  \label{exp-K-t}
\end{eqnarray}
where $\kappa$ is the hyperplane triple self-intersection, and
\begin{displaymath}
{\rm Li}_k(q) \: = \: \sum_{n=1}^{\infty} \frac{q^n}{n^k}, \: \: \:
q \: = \: \exp\left( 2 \pi i t \right).
\end{displaymath}
Now, one of the properties of
$t$ is that close to large radius, it should be defined up to a shift
by $1$, hence one expects
\begin{displaymath}
t \: = \: \frac{\ln z}{2 \pi i} \: + \: (\mbox{terms invariant under }z
\mapsto z e^{2 \pi i}).
\end{displaymath}
Hence, the correct normalization can be computed by dividing $Z_{nlsm}$
by the coefficient of $- (i/6)\kappa \ln(z \overline{z})^3/(2\pi i)^3$.
In the present case, $\kappa = 16$,
hence we should divide $Z_{nlsm}$ by
\begin{displaymath}
- i (2\pi i)^3 (z \overline{z})^{\mathsf{q}} f(0) \: = \:
- i (2\pi i)^3 (z \overline{z})^{\mathsf{q}}
\left| \sum_{k=0}^{\infty} z^k \frac{ \Gamma(1+2k)^4 }{ \Gamma(1+k)^8 } 
\right|^2
\end{displaymath}
to get
\begin{eqnarray}
\lefteqn{
\exp(-K) \: = \:
-i \frac{16}{6} \left[ \frac{\ln(z \overline{z})^3}{(2\pi i)^3} 
\: + \: \frac{8 \pi^2}{(2\pi i)^3} \frac{f'(0)}{f(0)}
\: - \: \frac{3}{2\pi i} \frac{\ln(z \overline{z})^2}{(2\pi i)^2} 
\frac{f'(0)}{f(0)} 
\right. } \label{exp-K-z-lr}\\
& & \hspace*{1.5in} \left.
\: - \: \frac{\ln(z \overline{z})}{2 \pi i} 
\left( \frac{8 \pi^2}{(2\pi i)^2} - \frac{3}{(2\pi i)^2} 
\frac{f''(0)}{f(0)} \right)
\: - \: \frac{1}{(2\pi i)^3} \frac{f^{(3)}(0)}{f(0)} \right].
\nonumber
\end{eqnarray}

Next, we need to solve for $t=t(z)$.  To do this, we compare the expression
above to equation~(\ref{exp-K-t}).

If we write 
\begin{equation}   \label{t-defn}
t(z) \: = \: \frac{\ln z}{2 \pi i} \: + \: \frac{\Delta(z)}{2\pi i}
\end{equation}
for some function $\Delta(z)$, then judging from the expression above,
the $\ln(z \overline{z})^2$ term (which should only arise in the
$(t - \overline{t})^3$ term in $\exp(-K)$) implies that
\begin{displaymath}
\frac{\Delta + \overline{\Delta}}{2\pi i} \: = \:
- \frac{1}{2 \pi i} \frac{ f'(0) }{f(0) }
\: = \:
- \frac{1}{2\pi i} \left. \frac{\partial}{\partial \epsilon} \ln f(\epsilon)
\right|_{\epsilon=0}.
\end{displaymath}
Define
\begin{displaymath}
g(\epsilon) \: = \: \sum_{k=0}^{\infty} z^k \frac{ \Gamma(1+2k - 2 \epsilon)^4}{
\Gamma(1 + k - \epsilon)^8}
\end{displaymath}
so that $f(\epsilon) = |g(\epsilon)|^2$, then
\begin{displaymath}
\Delta(z) \: = \: 2 \pi i C \: - \: \left.
\frac{\partial}{\partial \epsilon} \ln g(\epsilon) \right|_{\epsilon=0}
\end{displaymath}
for $C$ an undetermined real number, or equivalently
\begin{eqnarray*}
t(z) & = & \frac{\ln z}{2 \pi i} \: + \: C
\: - \: \frac{1}{2\pi i}
\left.
\frac{\partial}{\partial \epsilon} \ln g(\epsilon) \right|_{\epsilon=0}.
\end{eqnarray*}
Exponentiating equation~(\ref{t-defn}), we get
\begin{eqnarray*}
q & = &
\exp(2 \pi i t) \: = \: z e^{ 2 \pi i C} \exp\left( -
\left. \frac{\partial}{\partial \epsilon} \ln g(\epsilon) \right|_{\epsilon=0}
\right),
\\
& = &
z e^{2 \pi i C} \left( 1 \: + \: 64 z \: + \: 7072 z^2 \: + \:
991232 z^3 \: + \: 158784976 z^4 \: + \: 27706373120 z^5 
\right. \\
& & \hspace*{0.5in} \left.
\: + \: 5130309889536 z^6 \: + \: {\cal O}(z^7)
\right),
\end{eqnarray*}
which can be inverted to find 
\begin{eqnarray*}
z & = & q e^{-2 \pi i C} \: - \: 64 q^2 e^{-4 \pi i C}
\: + \: 1120 q^3 e^{-6 \pi i C} \: - \: 38912 q^4 e^{-8 \pi i C}
\: - \: 1536464 q^5 e^{-10 \pi i C} 
\\
& & \hspace*{0.5in} 
\: - \:
177833984 q^6 e^{-12 \pi i C}
\: - \: 19069001216 q^7 e^{-14 \pi i C} \: + \: {\cal O}(q^8).
\end{eqnarray*}

Next, we compare the expressions for $e^{-K}$ in equations~(\ref{exp-K-t}),
(\ref{exp-K-z-lr}).  In particular, equation~(\ref{exp-K-t}) contains
two different terms with Gromov-Witten invariants, each multiplied by a 
different power of $t$.  By demanding these two expressions match,
we should be able to get two independent expressions for the 
same Gromov-Witten invariants, which will provide a good consistency
check on our computations.

For definiteness, let us turn $z$'s into $t$'s, and compare coefficients
of various powers of $t$.  Applying equation~(\ref{t-defn}), 
we find that equation~(\ref{exp-K-z-lr}) can be rewritten in the form
\begin{eqnarray*}
e^{-K} & = & 
-i \frac{16}{6} \left[ (t - \overline{t})^3 \: + \:
\frac{ (t - \overline{t}) }{(2\pi i)^2} \left(
3 \left. \left( \frac{\partial}{\partial \epsilon} \right)^2 \ln f(\epsilon) 
\right|_{\epsilon=0}  \: - \: 8 \pi^2
\right) 
\right. \\
& & \hspace*{2.5in} \left.
\: - \: \frac{1}{(2\pi i)^3} \left.
\left( \frac{\partial}{\partial \epsilon} \right)^3
\ln f(\epsilon) \right|_{\epsilon=0} 
\right].
\end{eqnarray*}

Comparing the expression above to equation~(\ref{exp-K-t}),
we find that, from the coefficient of $(t - \overline{t})$,
\begin{displaymath}
- \frac{i}{(2 \pi i)^2} \sum_n n N_n \left( {\rm Li}_2(q^n)
\: + \: {\rm Li}_2(\overline{q}^n) \right) \: = \:
-i \frac{16}{6} \frac{1}{(2\pi i)^2} \left[ 
3 \left. \left( \frac{\partial}{\partial \epsilon} \right)^2 \ln f(\epsilon) 
\right|_{\epsilon=0} \: - \: 8 \pi^2
\right],
\end{displaymath}
and from the coefficient of $(1)$,
\begin{displaymath}
\frac{\zeta(3)}{4 \pi^3} \chi(X) \: + \:
\frac{2i}{(2\pi i)^3} \sum_n N_n \left( {\rm Li}_3(q^n) \: + \:
{\rm Li}_3(\overline{q}^n) \right) \: = \:
i \frac{16}{6} \frac{1}{(2\pi i)^3}
 \left.
\left( \frac{\partial}{\partial \epsilon} \right)^3
\ln f(\epsilon) \right|_{\epsilon=0}.
\end{displaymath}

This gives us two separate expressions for the Gromov-Witten invariants
$N_n$; by using both, we get a good consistency check of our results.
From a series expansion, each implies the (same) values, below, for
Gromov-Witten invariants:
\begin{eqnarray*}
N_1 & = & 512, \\
N_2 & = & 9728, \\
N_3 & = & 416256, \\
N_4 & = & 25703936, \\
N_5 & = & 1957983744, \\
N_6 & = & 170535923200,
\end{eqnarray*}
and constant $C = 0$.

Now, let us compare to existing results.
Counts of rational curves in ${\mathbb P}^7[2,2,2,2]$ are listed in
{\it e.g.} \cite{katz86}, \cite{hkty}[p. 36], which have the following 
results\footnote{
Reference \cite{katz86} lists the number of degree 1 curves as $512$,
and in \cite{hkty}[p. 36], we have been informed by one of the authors
(S.~Hosono) that rational curves of degree $d$ appear in the table as
the $n^r_{d+1}$ entry.  (Elliptic curves of degree $d$ are listed as
$n^e_d$.)
}
\begin{center}
\begin{tabular}{cc}
Degree & Count \\ \hline
1 & 512 \\
2 & 9728 \\
3 & 416256 \\
4 & 25703936 
\end{tabular}
\end{center}

Thus, we see that we have correctly computed the Gromov-Witten
invariants, a good consistency check of
this approach.

\section{Noncommutative resolution}
\label{sect:nc}

Now, let us turn to the Landau-Ginzburg phase of the same
GLSM, at $r \ll 0$.
Let us first derive an expression for the partition function,
and then proceed as above to derive analogues of Gromov-Witten invariants.
These will form our prediction for Gromov-Witten invariants of the
noncommutative resolution, and whether they match the Gromov-Witten
invariants of a smooth branched double cover will tell us whether
the SCFTs for the noncommutative resolutions may be continuously
connected by complex structure deformations to SCFT's for smooth
branched double covers.

Our analysis begins with the expression for the partition function
in equation~(\ref{start-part-fn}).  Since we are now considering the
$r \ll 0$ phase, we will close the contour on the right half-plane.
We will assume that $\mathsf{q}$ is just below $1/2$.

First, we will show that all of the poles of $Z_{\Phi}$ are cancelled
out by zeroes of the same.
From our previous analysis, recall that poles of the numerator of
$Z_{\Phi}$ are located at
\begin{displaymath}
\tau \: = \: m/2 \: - \: k
\end{displaymath}
for $k \geq 0$, and poles of the denominator are located at
\begin{displaymath}
\tau \: = \: 1 \: - \: m/2 \: + \: k_1
\end{displaymath}
for $k_1 \geq 0$.
When $m \geq 1$, the numerator can have poles inside the contour,
and this will happen for $k < m/2$.
If there is a pole inside the contour for some $k$,
then it is cancelled by a pole of the denominator with
\begin{displaymath}
k_1 \: = \: m \: - \: k \: - \: 1,
\end{displaymath}
which is guaranteed to be nonnegative by the fact that $k < m/2$ and
$m \geq 1$.

Now, let us count contributing poles of $Z_P$.
For one of the poles of $Z_P$ to lie inside the integration contour,
judging solely from the numerator, 
we should require $k \geq {\rm max}(0, -m)$.  However, some of those
poles will be cancelled by poles in the denominator of $Z_P$.
Suppose without loss of generality that $m < 0$.
If $k_1 = -2m - 1 - k$, then formally (ignoring signs on 
$k$, $k_1$) a pole in the
numerator and denominator will coincide.  To generate a pole inside the
contour, $k_1$ must satisfy $0 \leq k_1 < -m-1$, which is equivalent to
$-m \leq k < -2m$.  Hence, the only contributing poles will have
$k \geq {\rm max}(0, -2m)$, or equivalently $m \geq - k/2$.

Therefore, for $r \ll 0$, we can write the partition 
function in equation~(\ref{start-part-fn}) as
\begin{eqnarray*}
\lefteqn{
Z_{nlsm} \: = \:
- \sum_{k=0}^{\infty} \sum_{m \geq - k/2}
e^{-i m \theta} e^{-4 \pi \mathsf{q} r}
} \\
& & \hspace*{0.75in}
 {\rm Res}_{\tau = (1/2)(m+1+k)}
\left\{
e^{4 \pi r \tau} 
\left( \frac{ \Gamma( \tau - m/2 ) }{ \Gamma( 1- \tau - m/2) }
\right)^8
\left(
\frac{ \Gamma( 1 - 2 \tau + m ) }{ \Gamma(2 \tau + m ) }
\right)^4
\right\}, \\
& = & - \sum_{k=0}^{\infty} \sum_{m \geq - k/2} e^{- i m \theta}
e^{- 4 \pi \mathsf{q} r} \oint \frac{d\epsilon}{2 \pi i}
e^{2 \pi r(m + 1 + k)} e^{4 \pi r \epsilon} \\
& & \hspace*{0.75in} \cdot
\left(
\frac{ \Gamma( (k+1)/2 + \epsilon) }{
\Gamma( (1-k)/2 - m - \epsilon ) }
\right)^8
\left( 
\frac{ \Gamma( - k - 2 \epsilon ) }{
\Gamma( 2m + 1 + k + 2 \epsilon ) }
\right)^4,
\end{eqnarray*}
where the overall sign takes into account the orientation on the 
original $\tau$ contour.
Define $\ell = k + 2m$, then we can write 
\begin{eqnarray*}
Z_{nlsm} & = &
- \sum_{\delta = 0}^1 \sum_{k,\ell = \delta, 2+\delta, 4+\delta, \cdots}
e^{- i (\ell-k) \theta / 2} e^{- 4 \pi \mathsf{q} r} 
\oint \frac{d \epsilon}{2\pi i}
e^{2 \pi r ( \ell + k + 2)/2} e^{4 \pi r \epsilon} \\
& & \hspace*{0.75in} \cdot
\left(
\frac{ \Gamma( (k+1)/2 + \epsilon) }{
\Gamma( (1 - \ell)/2 - \epsilon) }
\right)^8
\left(
\frac{ \Gamma( -k - 2 \epsilon) }{
\Gamma( \ell + 1 + 2 \epsilon) }
\right)^4.
\end{eqnarray*} 
Define $a, b$ by
\begin{displaymath}
k \: = \: 2 a \: + \: \delta, \: \: \:
\ell \: = \: 2b \: + \: \delta,
\end{displaymath}
then
\begin{eqnarray*}
Z_{nlsm} & = &
- \sum_{\delta=0}^1 \sum_{a,b = 0}^{\infty}
e^{- i (b-a) \theta} e^{- 4 \pi \mathsf{q} r}
\oint \frac{d \epsilon}{2 \pi i} e^{2 \pi r (a + b + \delta + 1)}
e^{4 \pi r \epsilon} \\
& & \hspace*{0.75in} \cdot
\left(
\frac{ \Gamma( a + (\delta+1)/2 + \epsilon) }{
\Gamma( -b + (1 - \delta)/2 - \epsilon) }
\right)^8
\left(
\frac{ \Gamma( - 2a - \delta - 2 \epsilon) }{
\Gamma( 2b + \delta + 1 + 2 \epsilon) }
\right)^4.
\end{eqnarray*}
Define $z = \exp(- 2 \pi r + i \theta)$ as at large radius, then
\begin{eqnarray*}
Z_{nlsm} & = &
- \sum_{\delta=0}^1 \sum_{a,b = 0}^{\infty}
z^{-b} \overline{z}^{-a} 
\oint \frac{d \epsilon}{2 \pi i} (z \overline{z})^{\mathsf{q} - (\delta + 1)/2
- \epsilon} \\
& & \hspace*{0.75in} \cdot
\left( 
\frac{ \Gamma( a + (\delta+1)/2 + \epsilon) }{
\Gamma( -b + (1 - \delta)/2 - \epsilon) }
\right)^8
\left(
\frac{ \Gamma( - 2a - \delta - 2 \epsilon) }{
\Gamma( 2b + \delta + 1 + 2 \epsilon) }
\right)^4, \\
& = &
- \sum_{\delta=0}^1 \sum_{a,b = 0}^{\infty}
z^{-b} \overline{z}^{-a} 
\oint \frac{d \epsilon}{2 \pi i} (z \overline{z})^{\mathsf{q} - (\delta + 1)/2
- \epsilon} \pi^{-4} \\
& & \hspace*{0.5in} \cdot
\frac{ \left[ \sin\pi((\delta-1)/2 + \epsilon) \right]^8 }{
\left[ \sin \pi( \delta + 2 \epsilon) \right]^4 }
\frac{ \Gamma( a + (1 + \delta)/2 + \epsilon)^8 }{
\Gamma( 1 + 2a + \delta + 2 \epsilon)^4 }
\frac{ \Gamma(b + (1 + \delta)/2 + \epsilon)^8 }{
\Gamma( 1 + 2b + \delta + 2 \epsilon)^4 }.
\end{eqnarray*}

Proceeding as at large radius, define
\begin{displaymath}
f_{\delta}(\epsilon) \: = \: 
\left| 
\sum_{m=0}^{\infty} \left( \frac{1}{z} \right)^m 
\frac{ \Gamma(m + (1+\delta)/2 + \epsilon)^8 }{
\Gamma(2m + 1 + \delta + 2 \epsilon)^4 }
\right|^2,
\end{displaymath}
where the complex conjugation acts only on $z$.
Then we can write
\begin{eqnarray*}
Z_{nlsm} & = &
- \sum_{\delta=0}^1
\oint \frac{d \epsilon}{2 \pi i} (z \overline{z})^{\mathsf{q} - (\delta + 1)/2
- \epsilon} \pi^{-4}
\frac{ \left[ \sin\pi((\delta-1)/2 + \epsilon) \right]^8 }{
\left[ \sin \pi( \delta + 2 \epsilon) \right]^4 }
f_{\delta}(\epsilon),
\\
& = &
- (z \overline{z})^{\mathsf{q} - 1/2} \frac{1}{96 \pi^8}
\left(
- f_0(0) \ln(z \overline{z})^3 \: - \: 8 \pi^2 f_0'(0)
\: + \: 3 \ln(z \overline{z})^2 f_0'(0) 
\right. \\
& & \hspace*{1.25in} \left.
\: + \:
\ln(z \overline{z}) \left( 8 \pi^2 f_0(0) - 3 f_0''(0) \right)
\: + \: f_0^{(3)}(0) \right).
\end{eqnarray*}
The only contribution to the residue is from $\delta = 0$, which at
some level is a result of the fact that at large radius we have a 
complete intersection of quadrics.  For a more general case, 
one expects that some contributions from $\delta \neq 0$ might be nonzero,
which would impair our ability to apply \cite{jklmr2} to make
predictions for Gromov-Witten invariants.

To extract the mirror map, we need to find the triple self-intersection
$\kappa$.  For a smooth branched double cover, $\kappa = 2$, essentially
because it is a double cover of ${\bf P}^3$ -- $\kappa$ counts the 
number of elements in the cover, effectively.
In the present case, we want the analogue of $\kappa$ for a 
noncommutative resolution of a singular branched double cover.
We do not know how to define $\kappa$ in general for such; however,
the triple self-intersection can be computed away from the location of the
singularities, so we will assume $\kappa=2$ for the noncommutative
resolution also.

Proceeding as at large-radius, we should normalize $Z_{nlsm}$ so that
it matches $\exp(-K)$, which contains a 
\begin{displaymath}
- \frac{i}{6} \kappa (t - \overline{t})^3
\end{displaymath}
term.  As at large-radius, because of $B$ field shifts, $t$ should
have the form
\begin{displaymath}
t \: = \: \frac{\ln z}{2 \pi i} \: + \: (\mbox{terms invariant under }
z \mapsto z e^{2 \pi i} )
\end{displaymath}
and the term above should be the only possible source of a 
$\ln(z \overline{z})^3$ term.  Hence, the correct normalization should
be obtained by dividing $Z_{nlsm}$ by
\begin{displaymath}
\frac{ (z \overline{z})^{\mathsf{q} - 1/2} }{ (i/6) (2) }
\frac{1}{96 \pi^8} (-)  f_0(0) (2\pi i)^3,
\end{displaymath}
which yields
\begin{eqnarray*}
e^{-K} & = & 
- \frac{i}{6} (2) \left( 
\frac{\ln(z \overline{z})^3}{(2 \pi i)^3}
\: + \: \frac{8 \pi^2}{(2 \pi i)^3} \frac{f'_0(0)}{f_0(0)}
\: - \: \frac{3}{2 \pi i} \frac{ \ln(z \overline{z})^2 }{(2 \pi i)^2}
\frac{f'_0(0)}{f_0(0)}
\right. \\
& & \hspace*{1.25in} \left.
\: - \: \frac{ \ln(z \overline{z})}{2 \pi i}
\left( \frac{8 \pi^2}{(2 \pi i)^2} \: - \:
\frac{3}{(2\pi i)^2} \frac{f''_0(0)}{f_0(0)} \right)
\: - \: \frac{1}{(2\pi i)^3} \frac{f_0^{(3)}(0)}{f_0(0)} \right).
\end{eqnarray*}

If we write
\begin{displaymath}
t(z) \: = \: \frac{\ln z}{2 \pi i} \: + \: \frac{\Delta(z)}{2\pi i}
\end{displaymath}
then
\begin{displaymath}
(t - \overline{t})^3 \: = \:
\frac{\ln(z \overline{z})^3}{(2\pi i)^3} 
\: + \: 3 \left( \frac{\Delta + \overline{\Delta}}{2\pi i} \right)
\frac{ \ln(z \overline{z})^2
}{(2 \pi i)^2} \: + \: \cdots
\end{displaymath}
hence we read off that
\begin{eqnarray*}
\Delta + \overline{\Delta} & = & -  
\frac{f'_0(0)}{f_0(0)}
\\
& = & 
- \left. \frac{\partial}{\partial \epsilon} \ln g(\epsilon) 
\right|_{\epsilon=0} \: + \: {\rm c.c.},
\end{eqnarray*}
where
\begin{displaymath}
g(\epsilon) \: = \: \sum_{m=0}^{\infty} 
\left( \frac{1}{z} \right)^m 
\frac{ \Gamma(m + 1/2 + \epsilon)^8 }{\Gamma(2m + 1 + 2 \epsilon)^4}.
\end{displaymath}
This implies
\begin{displaymath}
q \: \equiv \: \exp(2 \pi i t) \: = \:
z e^{2 \pi i C} \exp\left( - 
\left. \frac{\partial}{\partial \epsilon} \ln g(\epsilon) \right|_{\epsilon=0}
\right),
\end{displaymath}
or more simply
\begin{eqnarray*}
q & = & 
z e^{2 \pi i C} \left(
65536 \: - \: 64 \frac{1}{z} \: - \:
\frac{93}{2048} \frac{1}{z^2} \: - \:
\frac{85}{1048576} \frac{1}{z^3} \: - \:
\frac{3251101}{17592186044416} \frac{1}{z^4}
\right. \\
& & \hspace*{0.65in} \left.
 \: - \:
\frac{8596595}{18014398509481984} \frac{1}{z^5}
\: + \: {\cal O}\left( \frac{1}{z^6} \right)
\right),
\end{eqnarray*}
where $C$ is an undetermined real constant.
Inverting, we find
\begin{eqnarray*}
\frac{1}{z} & = &
65536 q^{-1} e^{2 \pi i C} \: - \:
4194304 q^{-2} e^{4 \pi i C} \: + \:
73400320 q^{-3} e^{6 \pi i C} \: - \:
2550136832 q^{-4} e^{8 \pi i C}
\\
& & \hspace*{0.25in}
 \: - \:
100693704704 q^{-5} e^{10 \pi i C} \: - \:
11654527975424 q^{-6} e^{12 \pi i C} \: + \:
{\cal O}(q^{-7}).
\end{eqnarray*}

Now, proceeding as before, after algebra we can write
\begin{eqnarray*}
e^{-K} & = &
- \frac{i}{3} \left[ (t-\overline{t})^3 \: + \:
\frac{(t - \overline{t}) }{(2\pi i)^2}\left( 3
\left. \left( \frac{\partial}{\partial \epsilon} \right)^2
\ln f_0(\epsilon) \right|_{\epsilon=0} \: - \: 8 \pi^2 \right)
\right. \\
& & \hspace*{2.0in} \left.
\: - \: \frac{1}{(2\pi i)^3}
\left. \left( \frac{\partial}{\partial \epsilon} \right)^3 \ln f_0(\epsilon)
\right|_{\epsilon=0} \right].
\end{eqnarray*}
Comparing with equation~(\ref{exp-K-t}), we find that
\begin{displaymath}
\sum_n n N_n \left( {\rm Li}_2(q^{-n}) \: + \: {\rm Li}_2(\overline{q}^{-n})
\right) \: = \:
\frac{1}{3} \left(
 3
\left. \left( \frac{\partial}{\partial \epsilon} \right)^2
\ln f_0(\epsilon) \right|_{\epsilon=0} \: - \: 8 \pi^2 \right)
\end{displaymath}
and
\begin{displaymath}
\frac{ \zeta(3) \chi }{4 \pi^3} \frac{(2\pi i)^3}{i} \: + \:
2 \sum_n N_n \left( {\rm Li}_3(q^{-n}) \: + \: {\rm Li}_3(\overline{q}^{-n})
\right) \: = \:
\frac{1}{3}
\left. \left( \frac{\partial}{\partial \epsilon} \right)^3 \ln f_0(\epsilon)
\right|_{\epsilon=0}
\end{displaymath}
(where for obvious reasons we have replaced $q$ with $q^{-1}$).

By expanding each in series, one can extract predictions for Gromov-Witten
invariants, and doing so for both equations above gives a good consistency
check.  One finds, for both of the equations, that the Gromov-Witten 
invariants are given by
\begin{eqnarray*}
N_1 & = & 64, \\
N_2 & = &  1216, \\
N_3 & = & 52032, \\
N_4 & = & 3212992, \\
N_5 & = & 244747968, \\
N_6 & = & 21316990400, \\
N_7 & = & 2037544347200, \\
N_8 & = & 208507887048384, \\
N_9 & = & 22480719508041216,
\end{eqnarray*}
with constant $C=0$.

Now, let us compare to known results for generic smooth branched double
covers of ${\mathbb P}^3$.
Such smooth cases can be
described as hypersurfaces 
of the form
\begin{displaymath}
y^2 \: = \: f_8(x_1, \cdots, x_4),
\end{displaymath}
which is to say, 
${\mathbb P}^5_{[1,1,1,1,4]}[8]$,
and are discussed in {\it e.g.} \cite{davemirror};
table 3 in that reference lists 
\begin{center}
\begin{tabular}{cc}
Degree & Count \\ \hline
0 & 2 \\
1 & 29504 \\
2 & 128834912 \\
3 & 1423720545880 \\
4 & 23193056024793312
\end{tabular}
\end{center}

Thus, we see that the noncommutative resolutions of branched double
covers appearing in the GLSM for ${\mathbb P}^7[2,2,2,2]$, cannot be
continuously connected in SCFT moduli space by complex structure
deformations to smooth branched double covers, as the Gromov-Witten invariants
are demonstrably different.

Given this physics computation, it is also natural to ask to
what mathematics
this computation corresponds.  Given the mathematical definition of
noncommutative resolutions in terms of sheaf theory, it is not clear to
the author how one would go about directly defining Gromov-Witten
invariants mathematically -- perhaps these invariants are encoding
information about the CFT itself, rather than the noncommutative
structure {\it per se}.  However, there might be an indirect 
method\footnote{We would like to thank T.~Pantev for suggesting this,
and E.~Diaconescu for pointing out pertinent references.}.
Although a direct definition of Gromov-Witten invariants seems
obscure, direct definitions of Donaldson-Thomas invariants for such
noncommutative resolutions do exist
(see for example \cite{szendroi,nagao1,nagao2,toda}).
One might then be able to use the Donaldson-Thomas/Gromov-Witten
correspondence to formally define a set of integers, which would play
the same role as Gromov-Witten invariants, and might reasonably be
called Gromov-Witten invariants of a noncommutative resolution.
Perhaps the numbers we have computed could be obtained in this fashion.
We will leave such conjectured definitions to future work.

\section{Conclusions}

In this paper, we have applied the recent GLSM localization techniques
of \cite{jklmr2} to compute the Gromov-Witten invariants of an 
abstract CFT realizing a noncommutative resolution (in Kontsevich's sense)
of a singular branched double cover.  As those invariants do not match those
of related smooth branched double covers, we conclude that they cannot
be related by complex structure deformations in the (2,2) SCFT moduli
space.

It would also be interesting to apply the methods of \cite{jklmr2} to 
understand analogues of Gromov-Witten invariants for theories
close to Landau-Ginzburg orbifolds, as described in
{\it e.g.} \cite{fjr1,fjr2}.  In particular, invariants for the
Landau-Ginzburg point of the GLSM for the quintic in ${\mathbb P}^4$
were computed in \cite{c-r}, and it would be interesting to rederive
them using localization methods in GLSM's.  The methods described here
are not directly applicable:  for example,
the Landau-Ginzburg point of the quintic
does not have a $B$ field, so there is no analogue of the statement
that $t$ should contain a term proportional to $\ln z$, and indeed
the partition function at the Landau-Ginzburg point does not contain
terms involving $\ln z$'s.  Nevertheless, if a method could be found,
the derivation would be interesting.

\section{Acknowledgements}

We would like to thank D.-E. Diaconescu, S.~Hosono, H.~Jockers,
S.~Katz, T.~Pantev, M.~Romo, and especially J.~Lapan
for useful conversations.
We would also like to thank the IPMU in Tokyo, Japan, and the
Banff Center for hospitality while this work was completed.
This work was partially supported by NSF grant PHY-1068725.

\end{document}